\documentclass[twocolumn, prl,superscriptaddress]{revtex4-1}
\usepackage{graphicx}
\usepackage{epsfig}
\usepackage{color}
\usepackage{amsmath,bm}
\begin{document}
\title{$p$-wave contacts of quantum gases in quasi-one-dimensional and quasi-two-dimensional traps}
\author{Mingyuan He}
\affiliation{Shenzhen JL Computational Science and Applied Research Institute, Shenzhen, 518131, China}
\author{Qi Zhou}
\email{zhou753@purdue.edu}
\affiliation{Department of Physics and Astronomy, Purdue University, West Lafayette, IN, 47907, USA}
\affiliation{Purdue Quantum Science and Engineering Institute, Purdue University, West Lafayette, IN, 47907, USA}
\date{\today}
\begin{abstract}
The length scale separation in dilute quantum gases in quasi-one-dimensional or quasi-two-dimensional traps has spatially divided the system into two distinct regimes. Whereas universal relations defined in strict one or two dimensions apply in a scale that is much larger than the characteristic length of the transverse confinements, physical observables in the short distances are inevitably governed by three-dimensional contacts. Here, we show that $p$-wave contacts defined in different length scales are intrinsically connected by a universal relation, which depends on a simple geometric factor of the transverse confinements. While this universal relation is derived for one of the $p$-wave contacts, it establishes a concrete example of how dimensional crossover interplays with contacts and universal relations for arbitrary partial wave scatterings.
\end{abstract}

\maketitle

\section*{I. Introduction}
The study of contacts and universal relations has provided physicists with a powerful tool to explore the connections between two-body physics and many-body correlations in quantum gases and related systems. Since the first discovery by Shina Tan in 2005 \cite{Tan1,Tan2,Tan3}, contacts and universal relations have been generalized and applied to quantum systems for arbitrary partial wave scatterings in three-dimensional (3D) systems \cite{T4,N1,N2,N3,V1,V2,V6,P1,P2,P4,P5,Zhou1,Zhou2,Cui,Peng1} and have also been measured and verified in the experiments \cite{Vale,Jin1,Jin2,Jin3,P3,Vale2,Vale3}. By correlating different quantities through universal relations that are valid at any scattering lengths and any temperatures, contacts have been well accepted as the central quantity in dilute quantum systems. For instance, for $s$-wave scatterings, both the large momentum tail and the adiabatic relations, which concern how the energy changes with changing the scattering length, are controlled by the same contact. Meanwhile, the generalizations and applications to strictly one-dimensional (1D) \cite{Patu,V5,Cui1,Cui3,Sekino} and two-dimensional (2D) \cite{V3,V4,V7,Drut,Zhang1,Yu,George1} quantum systems have also been explored. However, strict 1D or 2D systems do not exist in reality and the transverse direction has a finite length scale. It has attracted considerable attention to study physics about the dimensional crossover in such systems \cite{Olshanii1,Ketterle,Petrov,Olshanii2,Qin,Cui2,Zhou3,Hu1,George2}.

\begin{figure}
	\centering
	{\includegraphics[width=0.47\textwidth]{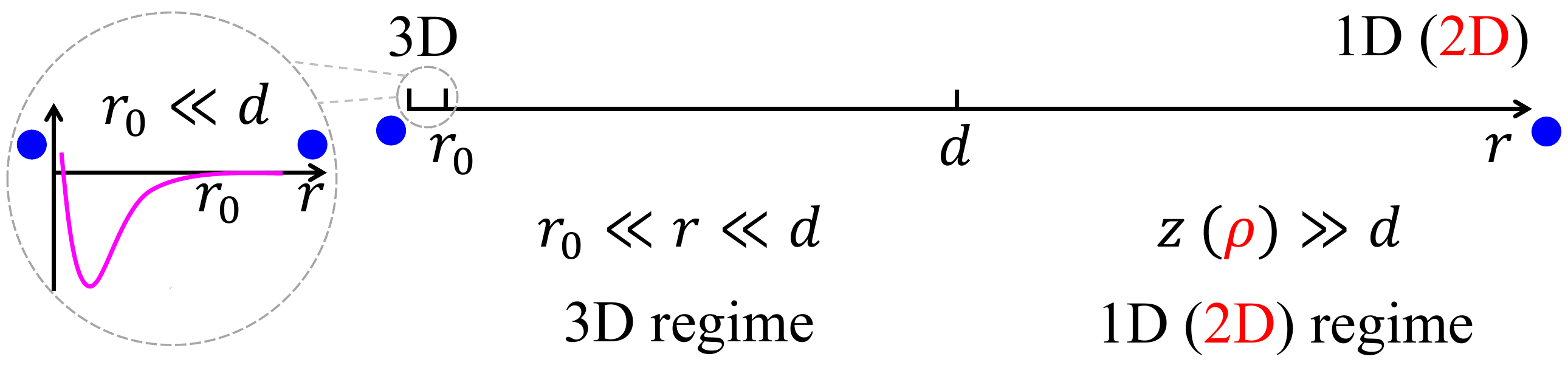}}
	\caption{The length scale separation in a quasi-1D (quasi-2D) trap. Blue spheres represent identical fermionic atoms. For a quasi-1D trap with strong harmonic confinement in the $x$ and $y$ directions, when the separation between two atoms is much larger (smaller) than $d$, two-body scatterings have 1D (3D) features and $\{C_{1D}\}$ ($\{C_{3D}\}$) controls all physical quantities in the corresponding large (small) length and small (large) momentum scales.  For a quasi-2D trap with a strong harmonic confinement along the $z$-direction (highlighted by red color), $\{C_{2D}\}$ ($\{C_{3D}\}$) controls the system in a scale $\rho\gg d$ ($r_0\ll r\ll d$). The Zoom in shows the short-range interaction between two atoms (purple curve) when $r\ll d$. }\label{Fig1}
\end{figure}

To explore physics in low dimensions, strong external confinements are often applied in the transverse directions. In such quasi-one-dimensional (quasi-1D) or quasi-two-dimensional (quasi-2D) traps, it is known that the two-body wavefunction behaves very differently in different spatial regimes. If we use $r_0$ to represent the range of the two-body interaction between atoms, a length scale separation exists in typical experiments on quantum gases, i.e., $r_0\ll d$, where $d$ is a characteristic length of the transverse confinements, as shown in Fig. \ref{Fig1}. For $s$-wave interactions, in a length scale that is much larger than $d$, the wavefunction takes the same form of that in strictly 1D (2D) systems. The universal relations derived in strictly 1D (2D) systems still work. $C_{1D}$ ($C_{2D}$), the corresponding 1D (2D) contact, controls other physical quantities. In contrast, at a short distance that is much smaller than $d$, the wavefunction recovers that in 3D. Universal relations established through $C_{3D}$ apply. Despite that $C_{1D}$ ($C_{2D}$) and $C_{3D}$ are defined in different length scales, it has been found that these contacts have intrinsic connections through a simple geometric factor determined only by $d$ \cite{V6,Zhou3}. 

In addition to $s$-wave scatterings, atoms could also interact with each other through $p$-wave and other high partial wave scatterings. For a single component Fermi gas, the $s$-wave scattering is suppressed due to the Pauli exclusion principle. Atom collisions are dominated by the p-wave scattering. Moreover, $p$-wave Feshbach resonances could be implemented to increase the $p$-wave scattering length \cite{Chin2010}. Other high partial wave scatterings have also been experimentally studied \cite{You2017}. A question arises naturally as to how $p$-wave and other high partial wave contacts manifest themselves in quasi-1D and quasi-2D traps. In particular, how high partial wave contacts in different length scales may correlate with each other? In this work, we focus on $p$-wave contacts and our results can be straightforwardly generalized to other high partial wave contacts. We show that the $p$-wave contact $C_{3D}$ can also be connected to $C_{1D}$ ($C_{2D}$) in quasi-1D (quasi-2D) traps through a characteristic geometric factor of quasi-1D (quasi-2D) traps. Moreover, this geometric factor is exactly as that obtained for $s$-wave contacts \cite{V6,Zhou3}. Since we concretize the discussions on $p$-wave contacts, the subscript to denote different partial waves has been suppressed. 
 
Some intrinsic differences with the $s$-wave scatterings include that the $p$-wave wavefunction is anisotropic and that multiple $p$-wave contacts are required. As such,  $\Psi({\bf r})\sim Y_{10}(\hat{\bf r})$ ($\Psi({\bf r})\sim Y_{1\pm1}(\hat{\bf r})$) in 3D regime and $\Psi({\bf r})\sim z/|z|$ ($\Psi({\bf r})\sim Y_{\pm1}(\hat {\bm \rho})$) in 1D (2D) regime. Consequently, compared with the results for $s$-wave scatterings, the first contact, $C_{3D}$ that determines the leading term of the large momentum tail, should be replaced by $C_{3D}|Y_{10}(\hat{\bf k}=\hat k_z)|^2$ ($C_{3D}|Y_{1\pm1}(\hat{\bf k}=\hat {\bf k}_\bot)|^2$) in quasi-1D (quasi-2D) traps for $p$-wave scatterings. $C_{2D}$ should be replaced by $C_{2D}|Y_{\pm1}(\hat {\bf k}_\bot)|^2$ in quasi-2D traps as well. As for the second contact $C_{3D}^1$ that determines the subleading term in the large momentum tail, the results are more elaborated. Unlike the first contact $C_{3D}$ (or $C_{1D}$, $C_{2D}$) that shows up in both the momentum distribution and the adiabatic relation, $C_{3D}^1$ (or $C_{1D}^1$, $C_{2D}^1$) is, in general, different from that in the adiabatic relation, unless the momentum of the center of mass of any pair of particles vanishes \cite{Zhang1,Cui1,Peng1}. To avoid this subtlety, we will consider the simplest situation in which the center of mass  does have a vanishing momentum. Under this situation, it turns out that $C_{3D}^1|Y_{10}(\hat{\bf k}=\hat k_z)|^2$ ($C_{3D}^1|Y_{1\pm1}(\hat{\bf k}=\hat {\bf k}_\bot)|^2$) can be connected to a combination of $C_{1D}$ and $C_{1D}^1$ ($C_{2D}|Y_{\pm1}(\hat {\bf k}_\bot)|^2$ and $C_{2D}^1|Y_{\pm1}(\hat {\bf k}_\bot)|^2$) through the same geometric factor as that for $C_{3D}$.

The main results of this paper are summarized as follows. (I) In quasi-1D (quasi-2D) traps, 1D (2D) $p$-wave contacts $\{C_{1D},C_{1D}^1\}$ ($\{C_{2D},C_{2D}^1\}$) should be defined from the momentum distribution $n_{\bf k}$ in the momentum regime, $k_F\ll k\ll d^{-1}$, where $k=|{\bf k}|$ and $k_F$ is the Fermi momentum. Specifically, by defining ${\bf k}=({\bf k}_\bot, k_z)$, we obtain
 \begin{equation}
n_{1D} (k_z)\equiv\int \frac{d{\bf k}_\bot}{(2\pi)^2} n_{\bf k}\stackrel{k_F\ll k_z\ll d^{-1}}{\xrightarrow{\hspace*{1.4cm}} } \frac{C_{1D}}{k_z^2}+\frac{C_{1D}^1}{k_z^4}, \label{nk1D}
\end{equation}
\begin{equation}
n_{2D}({\bf k}_\bot)\equiv\int  \frac{dk_z}{2\pi} n_{\bf k} \stackrel{k_F\ll k_\bot \ll d^{-1}}{\xrightarrow{\hspace*{1.4cm}} } (\frac{C_{2D}}{k_\bot^2}+\frac{C_{2D}^1}{k_\bot^4})|Y_1(\hat {\bf k}_\bot)|^2.
\label{2Dnk}
\end{equation}
Moreover, 3D $p$-wave contacts $\{C_{3D},C_{3D}^1\}$ govern $n_{\bf k}$ in the large momentum tail, $k\gg d^{-1}$, 
\begin{equation}
n_{\bf k}\stackrel{k\gg d^{-1}}{\xrightarrow{\hspace*{1cm}} }(\frac{C_{3D}}{k^2}+\frac{C_{3D}^1}{k^4})|Y_{1m}(\hat {\bf k})|^2.\label{nk3D}
\end{equation}
(II) We establish exact relations between $p$-wave contacts $\{C_{1D},C_{1D}^1\}$ ($\{C_{2D},C_{2D}^1\}$) and $\{C_{3D},C_{3D}^1\}$ in quasi-1D (quasi-2D) traps, which are
\begin{eqnarray}
&C_{3D}|Y_{10}(\hat{\bf k}=\hat k_z)|^2&=\pi d^2  C_{1D}, \label{C13}\\
&C_{3D}^1|Y_{10}(\hat{\bf k}=\hat k_z)|^2&=\pi d^2  (C_{1D}^1+\frac{4}{d^2} C_{1D}), \label{C131}
\end{eqnarray}
and
\begin{eqnarray}
&C_{3D}|Y_{1\pm1}(\hat{\bf k}=\hat {\bf k}_\bot)|^2&=\sqrt{\pi d^2}  C_{2D}|Y_{\pm1}(\hat{\bf k}_\bot)|^2, \label{C23}\\
&C_{3D}^1|Y_{1\pm1}(\hat{\bf k}=\hat {\bf k}_\bot)|^2&=\sqrt{\pi d^2}  (C_{2D}^1+\frac{2}{d^2} C_{2D})|Y_{\pm1}(\hat{\bf k}_\bot)|^2. \notag\\
&&\label{C231}
\end{eqnarray}
Therefore, there are two equivalent schemes to explore universal thermodynamic relations in quasi-1D (quasi-2D) traps for $p$-wave scatterings. One way is through $\{C_{3D},C_{3D}^1\}$ that control any physical systems, including highly anisotropic traps. The other way is using $\{C_{1D},C_{1D}^1\}$ ($\{C_{2D},C_{2D}^1\}$), which determine $n_{\bf k}$ in the intermediate momentum regime. (III) Based on Eqs. (\ref{C13}-\ref{C231}), we rigorously prove that the adiabatic relation derived for strictly 1D (2D) system is also exact in quasi-1D (quasi-2D ) traps.

The remaining parts of this paper are outlined as follows. In Sec. II, we consider single-component Fermi gases in quasi-1D traps. We define $p$-wave contacts in 3D and 1D regimes from the momentum distribution and establish the connections between contacts and universal relations derived in 3D and 1D regimes. Similar discussions for single-component Fermi gases in quasi-2D traps are given in Sec. III. Last but not least, we conclude our results in Sec. IV.

\section*{II. $p$-WAVE CONTACTS AND UNIVERSAL RELATIONS IN QUASI-1D TRAPS}
We consider single-component Fermi gases with short-range interactions between any two atoms such that only $p$-wave scatterings and $p$-wave contacts are relevant. We first consider $N$ identical fermionic atoms in a quasi-1D trap. The Hamiltonian is written as
\begin{equation}
H =  - \sum\limits_i {\frac{{{\hbar ^2}\nabla _i^2}}{{2M}}}  + \sum\limits_i V ({\rho _i}) + \sum\limits_{i > j} {U({\bf r}_{ij})} ,
\end{equation}
where $M$ is the atomic mass, ${\bf r}_i=({\bm \rho}_i, z_i)$ is the spatial coordinate of the $i$th atom, $\rho_i=|{\bm \rho}_i|$ and ${\bf r}_{ij}={\bf r}_i-{\bf r}_j$. $V(\rho_i)=\frac{1}{2}M\omega^2\rho^2_i$ is the harmonic trap of the $i$th atom in the $x$ and $y$ directions. Atoms are moving freely along the $z$-direction. $U({\bf r}_{ij})$ is a generic short-range interaction between the $i$th and $j$th atoms, which is finite only in the regime $|{\bf r}_{ij}|<r_0$. $V(\rho_i)$ is strong enough so that $d=\sqrt{2\hbar/(M\omega)}\ll k_F^{-1}$ is satisfied. It, in other words, means that the chemical potential $\mu$ is much smaller than the energy gap between the ground and the first vibration level of the harmonic trap, i.e., $\mu\ll 2\hbar\omega$. When the distance between any two atoms $r\ll k_F^{-1}$, the many-body wavefunction of the system has a universal asymptotic form \cite{N2,Zhou1,P4}
\begin{equation}
\Psi \stackrel{r\ll k_F^{-1}}{\xrightarrow{\hspace*{1cm}} }  \int d\epsilon_q \phi({\bf r};\epsilon_q)G(\frac{{\bf r}_1+{\bf r}_2}{2}, {\bf r}_{i\neq 1,2}; E-\epsilon_q), \label{Asym}
\end{equation}
where ${\bf r}={\bf r}_1-{\bf r}_2$ and $r=|{\bf r}|$. $\phi({\bf r};\epsilon_q)$ is the two-body relative wavefunction for $p$-wave scattering. $G(({{\bf r}_1+{\bf r}_2})/{2}, {\bf r}_{i\neq 1,2}; E-\epsilon_q)$ is the many-body wavefunction that includes the center of mass motion of the two atoms and the motion of all the other $N-2$ atoms. $\epsilon_q=\hbar \omega + \hbar^2q^2/M$ is the collision energy and $q$ is the corresponding momentum. $E$ is the total energy of the system. Eq. (\ref{Asym}) applies to any dilute quantum systems, regardless of the strength of the transverse confinement. The explicit form of the $p$-wave wavefunction $\phi({\bf r};\epsilon_q)$ at $r>r_0$ in quasi-1D traps is (See appendix A)
\begin{equation}
\begin{split}
\phi({\bf r};\epsilon_q) =& \Phi_{00}({\bm \rho})[\cot [{\eta_{1D}} (q)]\sin(q|z|)+\cos(q|z|)]\frac{z}{|z|}\\
&+\frac{z}{|z|}\sum_{n>0}\Phi_{n0}({\bm \rho})e^{-q_n |z|},\label{2bwf}
\end{split}
\end{equation}
where $\Phi_{nm}({\bm \rho})$ is the eigenstate of the 2D harmonic oscillator with eigenenergy $E_\bot^{nm}=\hbar\omega(2n+|m|+1)$, $n$ is the quantum number for the radial part of the wavefunction and $m$ is the angular momentum quantum number. As the axial direction of the quasi-1D trap, the $z$-direction, is chosen as the quantization axis, the relevant angular momentum number in 3D regime along this axis is $l=1$ and $m=0$, $\phi({\bf r};\epsilon_q) \sim Y_{10}(\hat{\bf r})\sim z/r$. For any fixed ${\bm \rho}=(x,y)$ in the transverse direction, the wavefunction is an odd wave, $\phi({\bf r};\epsilon_q)  \sim z/|z|$. When $|z|$ is fixed, the wavefunction in the $x$ and $y$ directions is proportional to $Y_{0}(\hat{\bm \rho})$. Due to the orthogonality of $Y_{m}(\hat{\bm \rho})=[(x+iy)/\rho]^{m}/\sqrt{2\pi}$, only the wavefunctions $\{\Phi_{nm}({\bm \rho})\}$ with $m=0$ are relevant. $\eta_{1D}(q)$ is the phase shift in 1D, which can be expanded in the low energy limit as $q\cot{\eta_{1D}(q)}=-1/a_{1D} + r_{1D}^e q^2$. $a_{1D}$ and $r_{1D}^e$ are the scattering length and effective range in 1D, respectively. The first line in Eq. (\ref{2bwf}) is contributed from the ground state of the harmonic trap, while the second line gives the contribution from the excited states. $q_n=\sqrt{(E_\bot^{n0}-\epsilon_q)M/\hbar^2}$. Typically, as $\hbar^2q^2/M \ll 2\hbar\omega$, $q_n$ is larger than zero for all positive integer $n$. Therefore, the second line in Eq. (\ref{2bwf}) decreases exponentially in 1D regime. By defining $z^*\equiv 1/q_1\sim d$, Eq. (\ref{2bwf}) reduces to the wavefunction in strict 1D when $|z|\gg z^*$. Clearly, $z^*\ll k_F^{-1}$ is satisfied. Moreover, by using the definition $n_{\bf k}=\sum\nolimits_{i=1}^{N} \int \prod\nolimits_{j\neq i} {d {\bf r}_j} \left| \int d{\bf r}_i \Psi e^{-i{\bf k}\cdot {\bf r}_i} \right|^2$, the momentum distribution of the many-body system in the regime $k_F\ll k\ll d^{-1}$ has the asymptotic form
\begin{equation}
n_{\bf k} \stackrel{k_F\ll k\ll  d^{-1}}{\xrightarrow{\hspace*{1cm}} } |\Phi_{00}({\bf k}_\bot)|^2(\frac{ C_{1D}}{k_z^{2}}+\frac{ C_{1D}^1}{k_z^{4}}), \label{nk1Dfull}
\end{equation}
where $\Phi_{00}({\bf k}_\bot)=\int d {\bm \rho} \Phi_{00}({\bm \rho}) e^{-i {\bf k_\bot} \cdot {\bm \rho}}$. The 1D $p$-wave contacts are defined as
\begin{eqnarray}
C_{1D}&=&4N(N-1)\int d{\bf R}_{12}\Big|g_{1D}^{(0)}\Big|^2, \label{C1D}\\
C_{1D}^1&=&4N(N-1)\int d{\bf R}_{12}\Big[g_{1D}^{(0)}g_{1D}^{(2)\ast}+g_{1D}^{(2)} g_{1D}^{(0)\ast}\Big], \label{C1D1}
\end{eqnarray}
where ${\bf R}_{12}$ denotes a coordinates set $\{({{\bf r}_1+{\bf r}_2})/{2}, {\bf r}_{i\neq 1,2}\}$ and $ d{\bf R}_{12}=\prod\nolimits_{i \ne 1,2} {{d}{{\bf r}_i}} d\left( {{{\bf r}_1} + {{\bf r}_2}} \right)/2$. $g_{1D}^{(s)}=\int d \epsilon_q q^s G ({\bf R}_{12};E-\epsilon_q)$. Since all the other momentum scales such as the center of mass momentum of a pair of atoms and the inverse of the scattering length are generally much smaller than $k_F$ in the strongly interacting regime, for simplicity, we have just specified that $k_F\ll k \ll d^{-1}$. In this regime, $n_{\bf k}$ is a broad distribution in the $k_x$ and $k_y$ directions, which is precisely a distinguishing feature of a quasi-1D system. When $k_F\ll k_z\ll d^{-1}$, the asymptotic form in Eq. (\ref{nk1Dfull}) could be extended to $k_{\bot}\rightarrow \infty$. By doing the integration over ${\bf k}_{\bot}$, we obtain Eq. (\ref{nk1D}).

When the distance between two atoms approaches the 3D regime with $r\ll d$, it is well-known that Eq. (\ref{Asym}) can be rewritten as
\begin{equation}
\Psi \stackrel{ r\ll d}{\xrightarrow{\hspace*{1cm}} } \int d\epsilon_q \phi_{3D}({\bf r};\epsilon_q) G_{3D}({\bf R}_{12};E-\epsilon_q), \label{Asym3D}
\end{equation}
where
\begin{equation}
\phi_{3D}({\bf r};\epsilon_q)\stackrel{ r\ll d}{\xrightarrow{\hspace*{1cm}} } (\frac{1}{r^2} + \frac{q_\epsilon^2}{2} + q_\epsilon^3 \cot \eta_{3D}\frac{r}{3}  ) Y_{1m}(\hat {\bf r}) \label{Asym3Dt}
\end{equation}
and $q_\epsilon=(M\epsilon_q/\hbar^2)^{1/2}$. $\eta_{3D}$ is the 3D phase shift, which can be expanded in the low energy limit as $q_\epsilon^3 \cot \eta_{3D}=-1/a_{3D} + r_{3D}^e q_\epsilon^2$. $a_{3D}$ and $r_{3D}^e$ are the scattering volume and effective range in 3D, respectively. Accordingly, $n_{\bf k}$ has a large momentum tail as shown in Eq. (\ref{nk3D}). The $p$-wave contacts in 3D regime are defined as
\begin{eqnarray}
C_{3D}&=&(4\pi)^2 N(N-1)\int d {\bf R}_{12} \Big|g_{3D}^{(0)}\Big|^2,\label{C3D} \\
C_{3D}^1&=&(4\pi)^2 N(N-1)\int d {\bf R}_{12} \Big[g_{3D}^{(0)}g_{3D}^{(2)\ast}+g_{3D}^{(2)}g_{3D}^{(0)\ast}\Big],\notag\\
&&\label{C3D1}
\end{eqnarray}
where $g_{3D}^{(s)}=\int d \epsilon_q q_\epsilon^s G_{3D} ({\bf R}_{12};E-\epsilon_q)$. Based on the method shown in appendix A, it is true that Eq. (\ref{2bwf}) can be written as
\begin{equation}
\begin{split}
\phi(0,z;\epsilon_q)\stackrel{ |z|\ll d}{\xrightarrow{\hspace*{0.5cm}} } \frac{d}{\sqrt{3}} \Big[ &  \frac{6}{d^2}(q\cot\eta_{1D} -\frac{2}{d} {\cal L}_0 + \frac{q^2d}{4} {\cal L}_1 ) \frac{|z|}{3} \\
&+ \frac{q_\epsilon^2}{2} + \frac{1}{z^2} \Big] Y_{10}(\hat {\bf r}=\hat z) \label{2bwf3D1D}
\end{split}
\end{equation}
when $|z| \ll d$ and $\rho=0$. Comparing it with Eq. (\ref{Asym3Dt}), one obtains $G_{3D}({\bf R}_{12};E-\epsilon_q)= (d/\sqrt{3}) G({\bf R}_{12};E-\epsilon_q)$ and
\begin{eqnarray}
\frac{1}{a_{1D}} & = & \frac{d^2}{6} (\frac{1}{a_{3D}} - \frac{2}{d^2} r_{3D}^e) - \frac{2}{d}{\cal L}_0, \label{sl13}\\
r_{1D}^e & = &  \frac{d^2}{6} r_{3D}^e -\frac{d}{4}{\cal L}_1, \label{er13}
\end{eqnarray}
where 
\begin{eqnarray}
{\cal L}_0 & = & \mathop {\lim }\limits_{\Lambda  \to \infty } \Big[ {\sum\limits_{n = 1}^\Lambda  {\sqrt n }  - \frac{2}{3}{\Lambda ^{3/2}} - \frac{1}{2}{\Lambda ^{1/2}}} \Big] =  - \frac{{\zeta (3/2)}}{{4\pi }}, \quad\\
{\cal L}_1 & = & \mathop {\lim }\limits_{\Lambda  \to \infty } \left[ {\sum\limits_{n = 1}^\Lambda  {\frac{1}{{\sqrt n }}}  - \int_0^\Lambda  {\frac{{dn'}}{{\sqrt {n'} }}} } \right] = \zeta (1/2).
\end{eqnarray}
$\zeta (s)=\sum\nolimits_{n=1}^\infty n^{-s}$ is the Riemann Zeta function. Equations (\ref{sl13}) and (\ref{er13}) are consistent with the results shown in reference \cite{Cui2}. Comparing Eq. (\ref{C1D}) and Eq. (\ref{C3D}), as well as Eq. (\ref{C1D1}) and Eq. (\ref{C3D1}), one immediately recognizes that Eqs. (\ref{C13}) and (\ref{C131}) hold.

\begin{figure}
	\centering
	\includegraphics[width=0.48\textwidth]{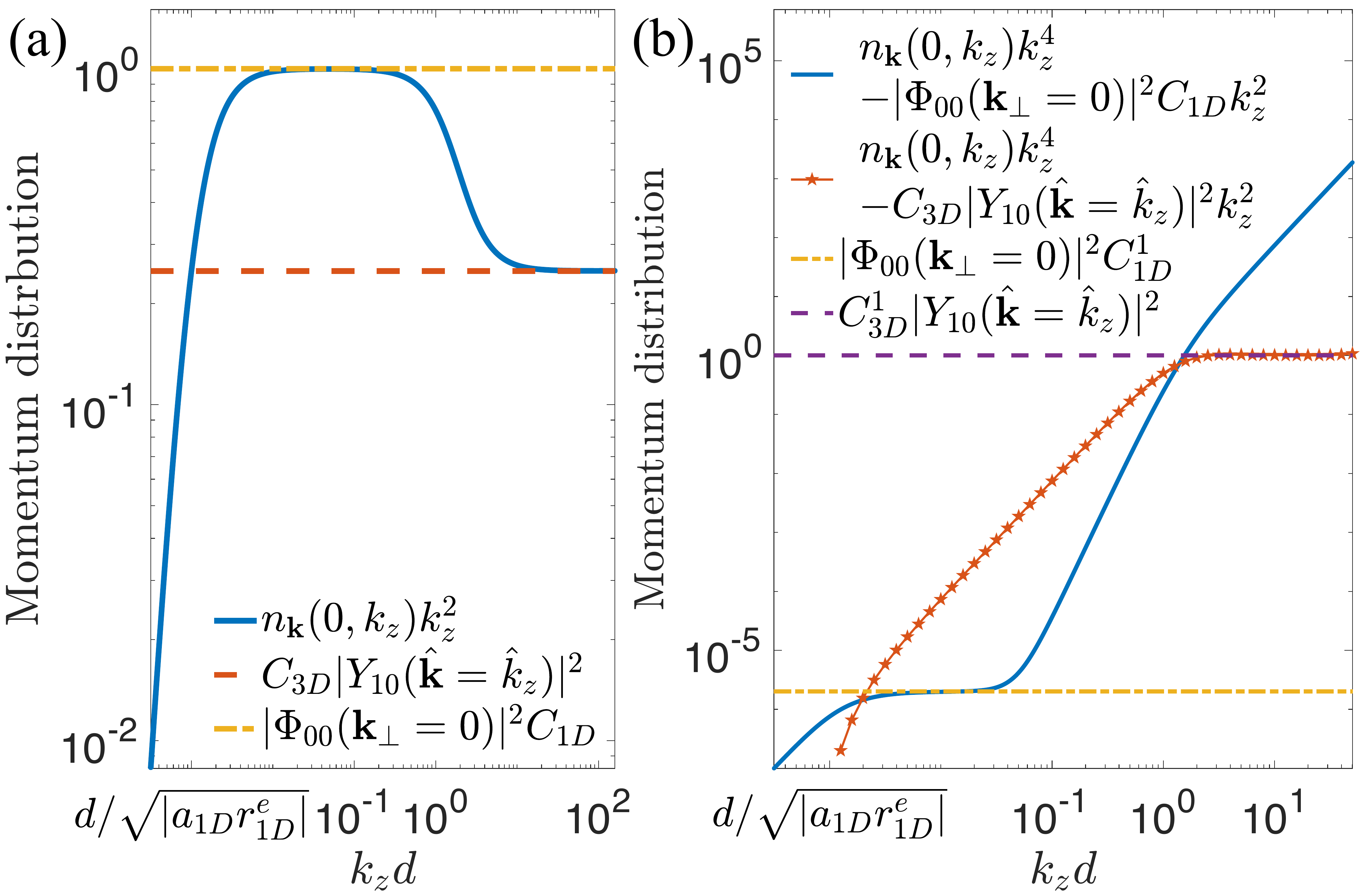}
	\caption{ (a) The scaled momentum $n_{\bf k}(0,k_z)k_z^2$ of a two-body system. It is determined by $C_{1D}$ and $C_{3D}$ in the regime $1/\sqrt{|a_{1D}r_{1D}^e|}\ll k_z\ll d^{-1}$ and $k_z\gg d^{-1}$, respectively. (b) Another scaled momentum $n_{\bf k}(0,k_z)k_z^4$ is determined by $C_{1D}^1$ and $C_{3D}^1$ in the regime $1/\sqrt{|a_{1D}r_{1D}^e|}\ll k_z\ll d^{-1}$ and $k_z\gg d^{-1}$, respectively. $n_{\bf k}$ is in unit of $d^2|\Phi_{00}({\bf k}_\bot=0)|^2 C_{1D}$. The total number of vibration levels considered is $N=300$, and $\sqrt{|a_{1D}r_{1D}^e|}=i/q=1000d$. }\label{Fig2}
\end{figure}

It turns out that the geometric factor $\pi d^2$ in Eqs. (\ref{C13}) and (\ref{C131}) is exactly as that obtained from the $s$-wave contacts. Whereas the quantitative result of the geometric factor is not obvious, a qualitative analysis could help explain why all partial wave scatterings may have the same geometric factor. We take the quasi-1D traps as an example. Whereas atoms are free along the $z$-direction, the 3D contact should be proportional to the 1D contact multiplied by the square of a length scale. It can be understood from the dimensional analysis, as the 3D and 1D $s$-wave contacts have the same dimension as $d^{-1}$ and $d^{-3}$, respectively. The cross-section area, $\pi d^2$, thus can naturally be the ratio between these two contacts. For high partial wave scatterings, the wavefunction is not isotropic in the real space. For instance, the $l$th partial wave wavefunction $\Psi \sim Y_{lm} (\hat{\bf r})$. The momentum distribution is then anisotropic and proportional to $|Y_{lm} (\hat {\bf k})|^2$. Thus, it is reasonable to expect that $C_{3D} |Y_{lm} (\hat{\bf k}=k_z)|^2$ is proportional to the 1D contact multiplied by the cross-sectional area $\pi d^2$. One can also write the universal relation for $s$-wave scatterings as $C_{3D} |Y_{00}(\hat{\bf k}=k_z)|^2 \sim  \pi d^2 C_{1D}$, where the Tan's definition of the $s$-wave contact is replaced by $C=C_{3D} |Y_{00}(\hat{\bf k})|^2$. To fix the exact numerical prefactor, however, calculations that we have presented here are necessary. The same argument applies to the relation between the second contacts $C_{3D}^1$ and $C_{1D}^1$ as well. The additional term of $C_{1D}$ in Eq. (\ref{C131}) is originated from the finite zero point energy of the ground state of the 2D harmonic oscillator. A similar argument applies to the quasi-2D traps.

Equation (\ref{C13}) provides us an explicit way to connect $n_{\bf k}$ in different momentum scales. Based on Eqs. (\ref{nk1D}) and (\ref{nk3D}), we have
\begin{equation}
\begin{split}
n_{\bf k} k^2\big|_{k\gg d^{-1},{\hat {\bf k}}=\hat k_z}=(\pi d^2)n_{1D}(k_z){k_z^2}\big|_{k_F\ll k_z\ll d^{-1}}.\label{n1D3D}
\end{split}
\end{equation}

To demonstrate the above results, we consider a two-body system and numerically solve for $n_{\bf k}$ using Eqs. (\ref{Asym}) and (\ref{2bwf}), which is basically just the Fourier transformation of Eq. (\ref{2bwf}). By counting on a sufficient number of excited states, we obtain $n_{\bf k}$, as shown in Fig. \ref{Fig2}. It clearly shows how the momentum distribution $n_{\bf k}(0,k_z)$ is changed from Eq. (\ref{nk1Dfull}) to Eq. (\ref{nk3D}) with the increase of $k_z$. The plateau of $n_{\bf k}(0,k_z)k_z^2$ in the regime $k_z\gg1/d$ ($|q|\ll k_z\ll 1/d$) in Fig. \ref{Fig2}(a) tells us the first 3D (1D) contact, $C_{3D}$ ($C_{1D}$). After obtaining $C_{3D}$ ($C_{1D}$), we can subtract its contribution to the momentum distribution and plot $[n_{\bf k}(0,k_z) -{ |Y_{10}(\hat{\bf k}=\hat k_z)|^2C_{3D}}/{k_z^{2}}]k_z^4$ ($[n_{\bf k}(0,k_z) -|\Phi_{00}({\bf k}_\bot=0)|^2{ C_{1D}}/{k_z^{2}}]k_z^4$). The plateau shown in Fig. \ref{Fig2}(b) then corresponds to the second 3D (1D) contact, $C_{3D}^1$ ($C_{1D}^1$).

In addition to $n_{\bf k}$, Eqs. (\ref{C13}) and (\ref{C131}) provide us a way to connect other universal thermodynamic relations in 1D and 3D as well. Here, we take the adiabatic relation as an example. In strictly 1D systems, where there is no transverse degree of freedom, the adiabatic relations for the odd-wave scatterings are written as \cite{Cui1}
\begin{eqnarray}
\frac{dE}{d(-1/a_{1D})} &=&\frac{\hbar^2C_{1D}}{4M}, \label{ab1D} \\
\frac{dE}{dr_{1D}^e} &=&\frac{\hbar^2C_{1D}^1}{8M}. \label{ab1D1} 
\end{eqnarray}
In quasi-1D systems, as aforementioned, a complete description of the system needs the introduction of not only $\{C_{3D},C_{3D}^1\}$ that capture physics in the length scale $z\ll d$ (or momentum scale $k\gg d^{-1}$), but also $\{C_{1D},C_{1D}^1\}$ that control physical quantities in a large length scale $z\gg d$ (or momentum scale $k\ll d^{-1}$). A natural question arises as to whether Eqs. (\ref{ab1D}) and (\ref{ab1D1}) are still valid. 

Interestingly, one can easily verify that Eqs. (\ref{ab1D}) and (\ref{ab1D1}) do hold for quasi-1D traps based on the following facts. (I) Since $\{C_{3D}, C_{3D}^1\}$ govern any 3D system, regardless of the shape and strength of the transverse confinements, the 3D adiabatic relations for $p$-wave scatterings \cite{P1,P2}
\begin{eqnarray}
\frac{dE}{d(-1/a_{3D})}&=&\frac{\hbar^2C_{3D}}{(4\pi)^2 2M },\label{ab3D}\\
\frac{dE}{dr_{3D}^e}&=&\frac{\hbar^2C_{3D}^1}{(4\pi)^2 4M}\label{ab3D1} 
\end{eqnarray}
are always valid in a quasi-1D trap. (II) Equations (\ref{C13}) and (\ref{C131}) establish exact relations between $\{C_{1D},C_{1D}^1\}$ and $\{C_{3D}, C_{3D}^1\}$. (III) As mentioned earlier, $\{a_{3D},r_{3D}^e\}$ and $\{a_{1D},r_{1D}^e\}$ are related by Eqs. (\ref{sl13}) and (\ref{er13}). Thus, by simply taking Eqs. (\ref{C13}) and (\ref{sl13}) into Eq. (\ref{ab3D}), Eq. (\ref{ab1D}) is obtained. As for Eq. (\ref{ab3D1}), the left hand side of which can be rewritten as
\begin{equation}
\frac{dE}{dr_{3D}^e}=\frac{dE}{dr_{1D}^e}\frac{dr_{1D}^e}{dr_{3D}^e}+\frac{dE}{d(-1/a_{1D})}\frac{d(-1/a_{1D})}{dr_{3D}^e}.
\end{equation}
Substituting Eqs. (\ref{C131}), (\ref{sl13}) and (\ref{er13}) into Eq. (\ref{ab3D1}), Eq. (\ref{ab1D1}) is obtained. Therefore, adiabatic relations derived for strictly 1D systems still valid for quasi-1D traps.

Though we concentrate on the adiabatic relation here, discussions apply to other universal thermodynamic relations as well. Thus, universal thermodynamic relations derived in 3D can be directly transformed to those derived in 1D through Eqs. (\ref{C13}) and (\ref{C131}) in quasi-1D traps.

\section*{III. $p$-WAVE CONTACTS AND UNIVERSAL RELATIONS IN QUASI-2D TRAPS}
In this section, we consider a quasi-2D trap. The Hamiltonian is written as
\begin{equation}
H =  - \sum\limits_i {\frac{{{\hbar ^2}\nabla _i^2}}{{2M}}}  + \sum\limits_i V (z_i) + \sum\limits_{i > j} {U({\bf r}_{ij})} ,
\end{equation}
where $V(z_i)=\frac{1}{2} M \omega^2 z^2_i$ is the harmonic trapping potential for the $i$th atom along the $z$-direction. $U({\bf r}_{ij})$ is a generic two-body interaction, which is finite only when $|{\bf r}_{ij}|<r_0$. Atoms are moving freely in the $x$ and $y$ directions. Essentially, the discussions are parallel to those for quasi-1D traps. As the length scale separation $r_0\ll k_F^{-1}$ still exists in quasi-2D traps, the asymptotic behavior of the many-body wavefunction, Eq. (\ref{Asym}), applies here as well. The explicit form of the $p$-wave wavefunction $\phi({\bf r};\epsilon_q)$ at $r>r_0$ in quasi-2D traps is (See appendix B)
\begin{equation}
\begin{split}
&\phi({\bf r};\epsilon_q) =  \Big\{\frac{\pi}{2} q\cot{\eta_{2D} }[J_1(q\rho)-\tan{\eta_{2D}} N_1(q\rho)]\Phi_{0}(z)\\
&-\frac{\pi}{2} \sum_{n>0}(-1)^n \sqrt{ \frac{(2n-1)!!}{(2n)!!} }q_n\Phi_{2n}(z)H_1^{(1)}(iq_{n}\rho)\Big\}Y_{\pm1}(\hat{\bm \rho}),\label{2bwf2D}
\end{split} 
\end{equation}
where $\eta_{2D}(q)$ is the 2D phase shift, which can be expanded in the low energy limit as $(\pi/2)q^2\cot{\eta_{2D}}-q^2\ln(qd/2)=-1/a_{2D}+r_{2D}^e q^2$ \cite{Randeria,Hammer2009,Hammer2010,Rakityansky}. $a_{2D}$ and $r_{2D}^e$ are the scattering area and effective range in 2D, respectively. $J_1$ and $N_1$ are the Bessel function of the first and second kind, respectively. $H_1^{(1)}=J_1+i N_1$. $\Phi_{n}(z)$ is the eigenfunction of the 1D harmonic oscillator with the corresponding eigen energy $E_z^n=\hbar\omega(n+1/2)$. $\epsilon_q=\hbar \omega/2 +\hbar^2q^2/M$. $q_n= \sqrt{ (E_z^{2n} -\epsilon_q )M/\hbar^2}$. By defining $\rho^*\equiv 1/q_1\sim d$, the wavefunction in Eq. (\ref{2bwf2D}) is 2D-like (3D-like) when $\rho>\rho^*$ ($\rho<\rho^*$). Again, by using the definition $n_{\bf k}=\sum\nolimits_{i=1}^{N} \int \prod\nolimits_{j\neq i} {d {\bf r}_j} \left| \int d{\bf r}_i \Psi e^{-i{\bf k}\cdot {\bf r}_i} \right|^2$, we can immediately obtain the tail of the momentum distribution. Equations (\ref{C23}) and (\ref{C231}) can then be obtained straightforwardly.

\begin{figure}
	\centering
	\includegraphics[width=0.48\textwidth]{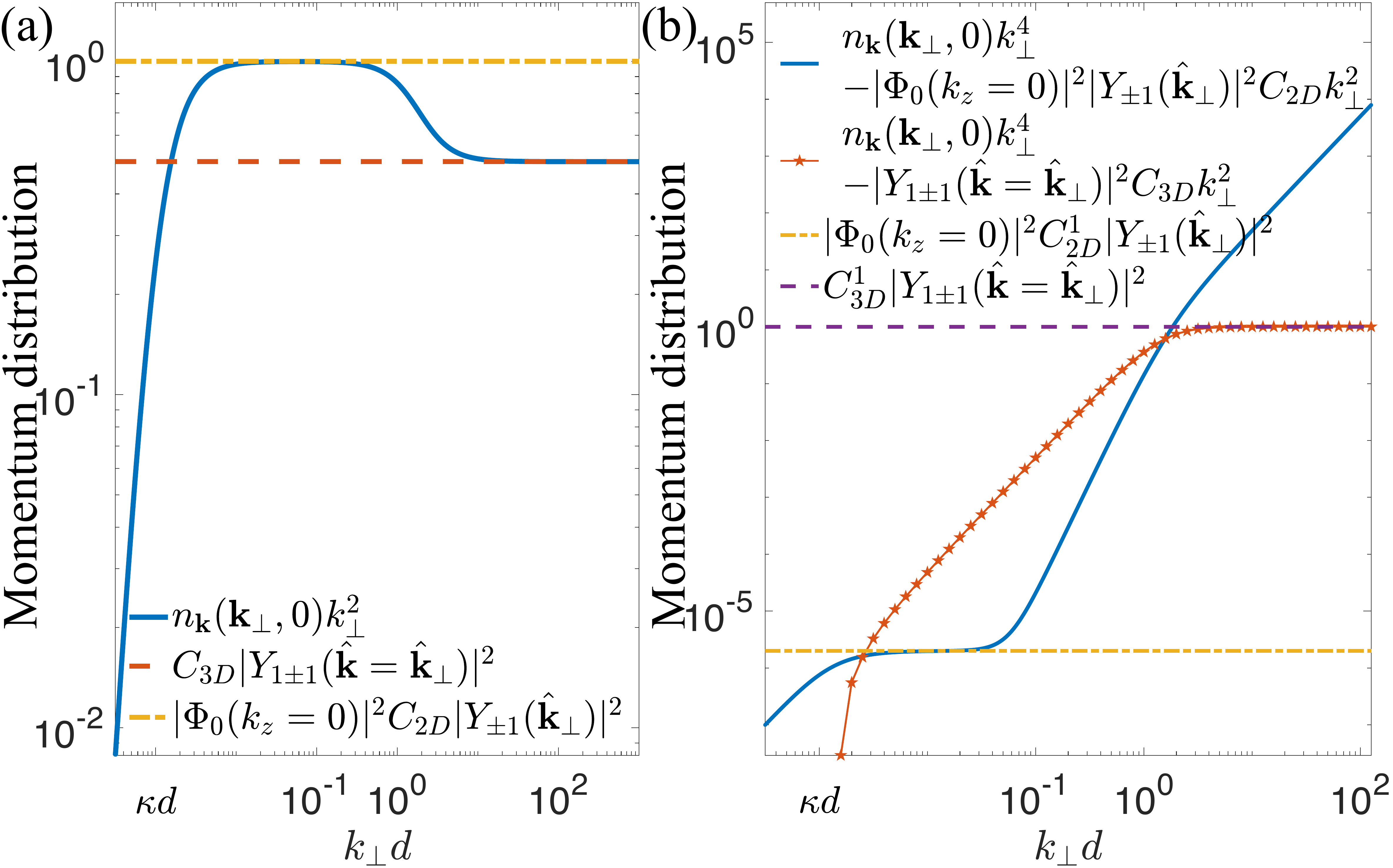}
	\caption{(a) The scaled momentum $n_{\bf k}({\bf k}_\bot,0)k_\bot^2$ of a two-body system. It is determined by $C_{2D}$ and $C_{3D}$ in the regime $\kappa\ll k_\bot\ll d^{-1}$ and $k_\bot\gg d^{-1}$, respectively. $\kappa=-iq \simeq 1/\sqrt{a_{2D}[(1/2)\ln(2a_{2D}/d^2)-r_{2D}^e]}$ \cite{Zhang1}. (b) Another scaled momentum $n_{\bf k}({\bf k}_\bot,0)k_\bot^4$ is determined by $C_{2D}^1$ and $C_{3D}^1$ in the regime $\kappa\ll k_\bot\ll d^{-1}$ and $k_\bot\gg d^{-1}$, respectively. $n_{\bf k}$ is in unit of $d^2|\Phi_{0} (k_z=0)|^2 C_{2D} |Y_{\pm1}(\hat{\bf k}_\bot)|^2$. The total number of vibration levels considered is $N=300$, and $\kappa=0.001/d$. }\label{Fig3}
\end{figure}

As a demonstration, Fig. \ref{Fig3} shows the numerical simulation for the momentum distribution of a two-body system, which is just the Fourier transformation of Eq. (\ref{2bwf2D}). Generally, when $k_F\ll k_\bot \ll d^{-1}$, we obtain
\begin{equation}
n_{\bf k} \stackrel{ k_F\ll k\ll  d^{-1}}{\xrightarrow{\hspace*{1cm}} } |\Phi_{0}(k_z)|^2(\frac{ C_{2D}}{k_\bot^{2}}+\frac{ C_{2D}^1}{k_\bot^{4}})|Y_{\pm1}(\hat {\bf k}_\bot)|^2,
\end{equation}
which shows a clear quasi-2D feature of $n_{\bf k}$ in this regime. By doing the integration over $k_z$, we obtain Eq. (\ref{2Dnk}). The 2D $p$-wave contacts are defined as
\begin{eqnarray}
C_{2D}&=&(2\pi)^2 N(N-1)\int d {\bf R}_{12} \Big|g_{2D}^{(0)}\Big|^2,\label{C2D} \\
C_{2D}^1&=&(2\pi)^2 N(N-1)\int d {\bf R}_{12} \Big[g_{2D}^{(0)}g_{2D}^{(2)\ast}+g_{2D}^{(2)}g_{2D}^{(0)\ast}\Big],\notag\\
&\label{C2D1}
\end{eqnarray}
where $g_{2D}^{(s)}=\int d\epsilon_q q^sG({\bf R}_{12},E-\epsilon_q)$. When the distance between two atoms approaches the 3D regime with $r\ll d$, Eqs. (\ref{Asym3D}-\ref{C3D1}) still apply in quasi-2D traps. Moreover, based on the method shown in appendix B, it turns out that $\phi({\bf r};\epsilon_q)$ has the asymptotic form the same as that shown in Eq. (\ref{Asym3Dt}) at $\rho \ll d$ and $z=0$. One obtains
\begin{equation}
\label{Asym3D2D}
\begin{split}
\phi({\bm \rho},0;\epsilon_q)  \stackrel{ \rho \ll d }{\xrightarrow{\hspace*{0.5cm}} }& \frac{\sqrt{d\sqrt{\pi}}}{\sqrt{3}} \Big\{ \frac{3}{d\sqrt{\pi}}\Big[\frac{\pi}{2}q^2\cot\eta_{2D} -q^2 \ln{\frac{qd}{2}} \\
&-\frac{4}{\pi d^2} W_1(\frac{q^2d^2}{4})+\frac{q^2}{\pi } W_2(\frac{q^2d^2}{4})\Big] \frac{\rho}{3} \\
&+ \frac{q_\epsilon^2}{2} + \frac{1}{\rho^2} \Big\}Y_{1\pm1}(\hat {\bf r}=\hat{\bm \rho}),
\end{split}
\end{equation}
where $q_\epsilon=(M\epsilon_q/\hbar^2)^{1/2}$,
\begin{equation}
\begin{split}
W_1(x) = &\mathop {\lim }\limits_{\Lambda  \to \infty } \Big[ \frac{\sqrt{\pi}}{9} \Lambda^{3/2} (3\ln \Lambda-2) + \frac{\sqrt{\pi}}{8} \sqrt{\Lambda} (\ln\Lambda +2) \\
&-\sqrt{\pi\Lambda} x - \frac{1}{2} \sum_{n=1}^{\Lambda} {B (n+\frac{1}{2},\frac{1}{2})n\ln{(n-x)}}\Big],
\end{split}
\end{equation}
\begin{equation}
\begin{split}
W_2(x) = \mathop {\lim }\limits_{\Lambda  \to \infty } \Big[ &\sqrt{\pi\Lambda} (\ln\Lambda -2) \\
&- \frac{1}{2} \sum_{n=1}^{\Lambda} {B (n+\frac{1}{2},\frac{1}{2})\ln{(n-x)}}\Big],
\end{split}
\end{equation}
and $B(x,y)=\Gamma(x)\Gamma(y)/\Gamma(x+y)$ is the beta function. Comparing Eq. (\ref{Asym3Dt}) with Eq. (\ref{Asym3D2D}), we obtain $G_{3D}({\bf R}_{12};E-\epsilon_q)= \sqrt{d\sqrt{\pi}/3} G({\bf R}_{12};E-\epsilon_q)$ and
\begin{eqnarray}
\frac{1}{a_{2D}} & = & \frac{\sqrt{\pi} d}{3} (\frac{1}{a_{3D}} - \frac{r_{3D}^e}{d^2}) - \frac{4}{d^2} \frac{W_1(0)}{\pi} ,\label{sl32}\\
r_{2D}^e & = & \frac{\sqrt{\pi} d}{3} r_{3D}^e + \frac{W_1^{(1)}(0)-W_2(0)}{\pi} ,\label{er32}
\end{eqnarray}
where $W_1(x)=W_1(0)+W_1^{(1)}(0)x +{\rm O} (x^2)$, $W_1(0)\approx 0.0853$, $W_1^{(1)}(0)=-\pi/2$ and $W_2(0)\approx (\pi/2)\ln[0.915/(2\pi)]$. Equations (\ref{sl32}) and (\ref{er32}) are consistent with the results shown in reference \cite{Zhang1}. Thus, the system is 3D-like when $r\ll d$ (or $k\gg d^{-1}$). $n_{\bf k}$ is controlled by $\{C_{3D},C_{3D}^1\}$ in this regime. Comparing Eqs. (\ref{C3D}) and (\ref{C3D1}) with Eqs. (\ref{C2D}) and (\ref{C2D1}), one can immediately see that Eqs. (\ref{C23}) and (\ref{C231}) hold. Similar to the relation shown in Eq. (\ref{n1D3D}), we can also make the connection that
\begin{equation}
n_{\bf k}k^2\big|_{k\gg d^{-1},\hat{\bf k}=\hat{\bf k}_\bot}=\sqrt{\pi d^2} n_{2D}({\bf k}_\bot){k_\bot^2}\big|_{k_F\ll k_\bot\ll d^{-1}}.
\end{equation}

In addition, we verify as well that the adiabatic relations for $p$-wave scatterings
\begin{eqnarray}
\frac{dE}{d(-1/a_{2D})} &=&\frac{\hbar^2C_{2D}}{(2\pi)^2 2M} ,\label{ab2D}\\
\frac{dE}{dr_{2D}^e}&=&\frac{\hbar^2C_{2D}^1}{(2\pi)^2 4M} ,\label{ab2D1}
\end{eqnarray}
which were originally derived for strictly 2D systems \cite{Zhang1}, still hold for quasi-2D traps. Substituting Eqs. (\ref{C23}), (\ref{C231}), (\ref{sl32}) and (\ref{er32}) into Eqs. (\ref{ab2D}) and (\ref{ab2D1}), the 3D adiabatic relations in Eqs. (\ref{ab3D}) and (\ref{ab3D1}) are recovered.

\section*{IV. CONCLUSION}
In conclusion, we have established exact relations between $p$-wave contacts, $\{C_{3D},C_{3D}^1\}$ and $\{C_{1D},C_{1D}^1\}$ ($\{C_{2D},C_{2D}^1\}$), in quasi-1D (quasi-2D) traps, which correlates physical quantities at different length or momentum scales, as well as universal relations in different dimensions. Despite that $p$-wave scatterings have very different properties compared to $s$-wave scatterings, the relations between 3D contacts and their counterparts in low dimensions are determined by the same geometric factor in quasi-1D (quasi-2D) traps.  Whereas the results of the first contacts, $C_{3D}$ and $C_{1D}$ ($C_{2D}$) are exact, the discussions about the second contacts, $C_{3D}^1$ and $C_{1D}^1$ ($C_{2D}^1$) have considered only the simplest case where the momentum of the center of mass of a pair of particles vanishes. It will be interesting to explore the effects of the motion of the center of mass in the future. We hope that our work will provide physicists a new way to study the dimensional crossover in quantum gases and related systems for arbitrary partial wave scatterings.

\section*{ACKNOWLEDGEMENTS}
M.H. is supported by HKRGC through HKUST3/CRF/13G and acknowledges the financial support from NSAF U1930402 and computational resources from the Beijing Computational Science Research Center. Q.Z. is supported by NSF PHY 1806796.  

\renewcommand{\theequation}{A\arabic{equation}}
\setcounter{equation}{0}
\section*{APPENDIX A: DERIVATION OF EQUATIONS (\ref{2bwf}) AND (\ref{2bwf3D1D})}
The explicit form of the $p$-wave wavefunction $\phi({\bf r};\epsilon_q)$ at $r>r_0$ in quasi-1D traps can be obtained by extending the wavefunction to the origin and solving the Schr\"odinger equation with $U({\bf r})$ replaced by the Huang-Yang pseudopotential \cite{HY1,HY2}. The Huang-Yang pseudopotential is introduced based on the fact that
\begin{equation}
\phi({\bf r};\epsilon_q)\sim [j_l(q_\epsilon r) -\tan \eta_{3D} n_l (q_\epsilon r)]Y_{lm}(\hat{\bf r})
\end{equation}
when $r\ll d$, and
\begin{eqnarray}
\nabla^2 j_l (qr)Y_{lm}(\hat{\bf r}) & = & -q^2j_l (qr)Y_{lm}(\hat{\bf r}), \\
\nabla^2 n_l (qr)Y_{lm}(\hat{\bf r}) & = & \left[-q^2n_l (qr)+ \frac{(2l+1)!!}{q^{l+1}} \frac{\delta(r)}{r^{l+2}}\right]Y_{lm}(\hat{\bf r}), \notag\\
&&
\end{eqnarray}
where $j_l$ ($n_l$) is the spherical Bessel function of the first (second) kind and $Y_{lm}(\hat{\bf r})$ is the spherical harmonics in 3D. One has
\begin{equation}
\begin{split}
\epsilon_q \phi({\bf r};\epsilon_q)=&\Big[-\frac{\hbar^2}{M} \nabla_\rho^2 +\frac{1}{4}M\omega^2 \rho^2 -\frac{\hbar^2}{M} \frac{\partial^2}{\partial z^2} \\
&-  \frac{\hbar^2}{2M} \frac{\tan \eta_{3D}}{q_\epsilon^3}\frac{3\delta(r)}{r^3} \Big(\frac{\partial}{\partial r}\Big)^3 (r^2\cdot)\Big] \phi({\bf r};\epsilon_q). \label{AQ1D}
\end{split} 
\end{equation}
By expanding $\phi({\bf r};\epsilon_q)$ in the basis $\{\Phi_{n0}({\bm \rho})\}$, one has
\begin{equation}
\phi({\bf r};\epsilon_q)=R_0 (z) \Phi_{00}({\bm \rho})+\sum_{n=1}^\infty a_n R_n (z) \Phi_{n0}({\bm \rho}).
\end{equation}
By taking it back to Eq. (\ref{AQ1D}) and considering a finite $r$, one has
\begin{eqnarray}
R_0 (z) & =& [\cot [{\eta_{1D}} (q)]\sin(q|z|)+\cos(q|z|)]\frac{z}{|z|},  \\
R_{n>0}(z) & =&e^{-q_n |z|} \frac{z}{|z|}.
\end{eqnarray}
$a_n$ is determined by using the boundary condition at $r=0$. Based on the fact that $\partial_z (z/|z|) = 2\delta(z)$, $\partial_z \delta(z) =-\delta (z)/z$ and $\delta (r)/r^2 = 4\pi \delta({\bf r})=4\pi \delta({\bm \rho})\delta(z)$, one has
\begin{equation}
\Phi_{00}({\bm \rho}) +\sum_{n=1}^\infty a_n \Phi_{n0}({\bm \rho}) \propto \delta ({\bm\rho}),
\end{equation}
which gives that $a_{n>0}=1$. Equation (\ref{2bwf}) is then obtained. Equation (\ref{2bwf3D1D}) can be obtained by using a similar method used in reference \cite{Olshanii1}. By introducing a $Q$ function as
\begin{equation}
\begin{split}
Q_n(a,b)&=\int_{n-1}^n e^{-a\sqrt{n'-b}}dn'\\
=&\int_{n-1}^n e^{-a\sqrt{n'}}dn'+\frac{ab}{2} \int_{n-1}^n \frac{e^{-a\sqrt{n'}}}{\sqrt{n'}}dn'+{\rm O} (b^2),
\end{split}
\end{equation}
one can write the summation in Eq. (\ref{2bwf}) as
\begin{equation}
\begin{split}
\sum_{n=1}^\infty e^{-(2|z|/d)\sqrt{n-b}}&=\sum_{n=1}^\infty \Big[e^{-(2|z|/d)\sqrt{n-b}}-Q_n(\frac{2|z|}{d},b)\\
&+Q_n(\frac{2|z|}{d},b)\Big]  -\frac{1}{2} \mathop {\lim }\limits_{\Lambda  \to \infty } e^{-(2|z|/d)\sqrt{\Lambda}}.
\end{split}
\end{equation}
The third term on the right hand side of the above equation can be integrated out directly. By doing the Talyor expansion with respect to $|z|$ to the rest terms, equation (\ref{2bwf3D1D}) can then be obtained.

\renewcommand{\theequation}{B\arabic{equation}}
\setcounter{equation}{0}
\section*{APPENDIX B: DERIVATION OF EQUATIONS (\ref{2bwf2D}) AND (\ref{Asym3D2D})}
Similar to the discussions in Appendix A, the explicit form of $\phi({\bf r};\epsilon_q)$ at $r>r_0$ in quasi-2D traps can be obtained by solving the Schr\"odinger equation
\begin{equation}
\begin{split}
\epsilon_q \phi({\bf r};\epsilon_q)=&\Big[ -\frac{\hbar^2}{M} \frac{\partial^2}{\partial z^2} +\frac{1}{4}M\omega^2 z^2  -\frac{\hbar^2}{M} \nabla_\rho^2\\
&-  \frac{\hbar^2}{2M} \frac{\tan \eta_{3D}}{q_\epsilon^3}\frac{3\delta(r)}{r^3} \Big(\frac{\partial}{\partial r}\Big)^3 (r^2\cdot)\Big] \phi({\bf r};\epsilon_q). \label{AQ2D}
\end{split} 
\end{equation}
By expanding $\phi({\bf r};\epsilon_q)$ in the basis $\{\Phi_{2n}(z)\}$ with even parity, one has
\begin{equation}
\phi({\bf r};\epsilon_q)=R_0({\bm \rho}) \Phi_{0}(z)+\sum_{n=1}^\infty a_{n} R_{n}({\bm \rho}) \Phi_{2n}(z).
\end{equation}
By taking it back to Eq. (\ref{AQ2D}) and considering the finite $r$, one has
\begin{eqnarray}
R_0 ({\bm \rho}) & =& \frac{\pi}{2} q\cot{\eta_{2D} }[J_1(q\rho)-\tan{\eta_{2D}} N_1(q\rho)]Y_{\pm1}(\hat{\bm \rho}),  \notag\\
&&\\
R_{n>0}({\bm \rho}) & =&-\frac{\pi}{2} q_{n}H_1^{(1)}(iq_{n}\rho)Y_{\pm1}(\hat{\bm \rho}),
\end{eqnarray}
where $Y_{m}(\hat{\bm \rho})=[(x+iy)/\rho]^{m}/\sqrt{2\pi}$ is the generalized spherical harmonics in 2D. $a_n$ is determined by using the boundary condition at $r=0$. Based on the fact that $\delta(\rho)/\rho =2\pi \delta({\bm \rho})$ and
\begin{eqnarray}
\nabla_\rho^2 J_l (q\rho)Y_{l}(\hat{\bm \rho}) & = & -q^2J_l (q\rho)Y_{l}(\hat{\bm \rho}), \\
\nabla_\rho^2 N_l (q\rho)Y_{l}(\hat{\bm \rho}) & = & \left[-q^2N_l (q\rho)+\frac{2}{\pi} \frac{(2l)!!}{q^l} \frac{\delta(\rho)}{\rho^{l+1}}\right]Y_{l}(\hat{\bm \rho}), \notag\\
&&
\end{eqnarray}
one has
\begin{equation}
\Phi_{0}(z) +\sum_{n=1}^\infty a_n \Phi_{2n}(z) \propto \delta (z),
\end{equation}
which gives that $a_{n>0}=(-1)^n\sqrt{[(2n-1)!!/(2n)!!]}$. Equation (\ref{2bwf2D}) is then obtained. Equation (\ref{Asym3D2D}) can also be obtained by using a similar method used in reference \cite{Olshanii1}. By introducing a $Q$ function as
\begin{equation}
\begin{split}
Q_n(\rho,q_\rho)=&\left[\frac{2}{\rho^2}+\frac{d^{-2}-2q_\rho^2}{2}(\gamma -\frac{1}{2})+\frac{1}{d^2} -\frac{q_\rho^2}{2}\right]\\
&\cdot \frac{\sqrt{\pi}\rho}{2}\int_{n-1}^n\frac{e^{-\frac{\rho^2n'}{\lambda_0^2d^2}}}{\sqrt{n'}}dn' \\
&+(\gamma-\frac{1}{2}+\frac{1}{2\lambda_0^2})\frac{\sqrt{4\pi}\rho}{d^2}\int_{n-1}^n \sqrt{n'}e^{-\frac{\rho^2n'}{\lambda_1^2d^2}}dn'\\
&+\frac{d^{-2}-2q_\rho^2}{4} \frac{\sqrt{\pi}\rho}{2} \int_{n-1}^n\frac{e^{-\frac{\rho^2n'}{\lambda_2^2d^2}}}{\sqrt{n'}}\ln\frac{\rho^2n'}{d^2}dn' \\
&+ \frac{\sqrt{\pi} \rho}{d^2} \int_{n-1}^n\sqrt{n'}e^{-\frac{\rho^2n'}{\lambda_3^2d^2}}\ln\frac{\rho^2n'}{d^2}dn'\\
&+ \frac{3\pi \lambda_0\lambda_4}{8d} \mathop {\lim }\limits_{\Lambda  \to \infty } \exp{(-\rho^3\Lambda^{5/2}/d^3)},
\end{split}
\end{equation}
where $\gamma$ is the Euler's constant, 
\begin{eqnarray}
\lambda_0 & = & \sqrt{\pi}, \\
\lambda_0 & = & \frac{\lambda_2[\gamma+\ln(4/\lambda_2^2)]}{2\gamma+\lambda_4+1},\\
\lambda_1^3 &=& \frac{\lambda_3^3[\gamma-2+\ln(4/\lambda_3^2)]-\sqrt{\pi}}{2\gamma+1/\pi-1},\\
{\rm Re}(\lambda_{1,2,3}^2)&>&0,
\end{eqnarray}
one can then write the summation term in Eq. (\ref{2bwf2D}) as
\begin{equation}
\begin{split}
&\sum_{n=1}^\infty\left[-\frac{\pi}{2}\frac{(2n-1)!!}{(2n)!!} q_n H_1^{(1)} (iq_n\rho) \right] \\
=&\sum_{n=1}^\infty\left[-\frac{\pi}{2}\frac{(2n-1)!!}{(2n)!!} q_n H_1^{(1)} (iq_n\rho) \right.\\
&\left.\qquad -\frac{Q_n(\rho,q_\rho)}{\pi} + \frac{Q_n(\rho,q_\rho)}{\pi}\right].
\end{split}
\end{equation}
Equation (\ref{Asym3D2D}) can then be obtained.

\end{document}